\documentclass[a4paper]{article}

\begin {document}

\title {\bf Quantum Law of Motion: Analysis and Extension 
to Higher Dimensions }
\author{A.~Bouda\footnote{Electronic address: 
{\tt bouda\_a@yahoo.fr}} 
\ and A.~Gharbi\footnote{Electronic address: 
{\tt hakimgharbi@yahoo.fr}}\\
Laboratoire de Physique Th\'eorique, Universit\'e de B\'eja\"\i a,\\ 
Route Targa Ouazemour, 06000 B\'eja\"\i a, Algeria\\}

\date{\today}

\maketitle
  
\begin{abstract}
\noindent 
In this paper, we review the recently formulated quantum laws
of motion and provide new observations. We also extend these  
laws to higher dimensions. By applying in two dimensions the obtained 
relations to charge submitted to an electric central potential, 
we decide between these laws. 
Furthermore, we extend the selected law to the relativistic case in 
higher dimensions. 

\end{abstract}

\vskip\baselineskip



\newpage

\section{Introduction}

Recently, Matone suggested that the quantum potential generates 
the gravitational potential \cite{M}. This hypothesis was the logical 
result of the quantum mechanics formulation based on the equivalence 
postulate \cite{FM1,FM2}. In this formulation, it is the 
quantum Hamilton-Jacobi equation (QHJE) which is investigated. 
In this context, it is shown that tunnel effect, energy 
quantization \cite{FM2,FM3} and band structure for the energy 
spectrum in the Kr\"onig-Penney model \cite{BM1} directly follow from 
the QHJE  without appealing to the usual axiomatic interpretation of 
the wave function. Another interesting feature of the QHJE is the fact 
that the quantum potential, which represents an additional term 
compared with the classical Hamilton-Jacobi equation (CHJE), can be 
seen as a term which guarantees the invariance of the Hamilton-Jacobi 
equation under any coordinate transformation \cite{FM2}. This reminds us 
of the role played by  the gravitational field in general relativity. In this 
paper, we will review the different quantum laws of motion based on the 
QHJE. Among them, we can cite the Bohm approach \cite{Bohm} for which 
Einstein raised  a serious objection in the case where the system is 
described by a real wave function. Einstein's criticism was 
answered by taking the wave function in the form \cite{FM1,FM2,B1}
\begin {equation}
\phi=\textsc{R}\left[\alpha \exp\left(i{S_0\over \hbar}\right)
+ \beta \exp\left(-i{S_0\over \hbar}\right)\right] ,
\end {equation}
where $\alpha$ and $\beta$ are complex constants.
We can also cite Floyd's proposal for which trajectories are 
obtained by using Jacobi's theorem \cite{Fl1}. However, in 
this formulation, it seems that there is some confusion in the 
definition of time parametrization \cite {BD1, BD2, BD3}. 
We essentially devote our discussion to the quantum laws 
established in \cite{BD1} and \cite{B2}.

The paper is organized as follows. In Section 2, we 
review some features of the Quantum Laws of motion 
proposed in \cite{BD1} and \cite{B2}. In Section 3, we extend these 
laws to higher dimensions. In Section 4, we apply in two dimensions 
the obtained relations to hydrogen atom and then decide between them. 
In Section 5, we extend the selected law to the relativistic case in 
higher dimensions. Section 6 is devoted to conclusion.

\section{The Quantum Laws of Motion}

From the one-dimensional stationary QHJE 
\begin{eqnarray}
{1\over2m}{\left(\partial{S_0}\over\partial{x}\right)^2}+
V\left(x\right)-E=\hskip60mm&& \nonumber \\
{\hbar ^2\over4m}\left[{{3\over2}
\left(\partial{S_0}\over{\partial{x}}\right)^{-2}
\left(\partial^2{S_0}\over\partial{x^2}\right)^2}-
{\left(\partial{S_0}\over\partial{x}\right)^{-1}}
{\left(\partial^3{S_0}\over\partial{x^3}\right)}\right],
\end{eqnarray}
and by appealing to the coordinate transformation \cite{FM2,FM4}
\begin {equation}
x \to \hat {x}  \ \ / \ \
{\partial\hat{x}  \over \partial x }={\partial S_0/ \partial x \over
 \sqrt{2m\left(E-V(x)\right)}} \; ,
\end {equation}
after which the above QHJE takes the classical form, 
\begin {equation}
{1\over 2m}{\left(\partial \hat {S}_0(\hat{x})\over 
\partial \hat{x}\right)}^2+\hat{V}(\hat{x})=\hat{E}  \; ,
\end {equation}
it is established in \cite{BD1} that
\begin {equation}
{1\over 2}{\partial {S}_0 \over \partial {x}} \; \dot{x} +V(x)=E \; .
\end {equation}
This law, with the use of (2), has allowed to establish the Quantum 
Newton Law \cite{BD1}. We first observe that relation (5) 
itself constitutes a law of motion. In fact, by using in (5) 
the solution \cite{B1,BD1,BD3} of (2)
\begin {equation}
S_0=
{\hbar} \arctan{\left[a {{\phi_1} \over
{\phi_2}}+b \right]}+{\hbar}{\lambda} \; ,
\end {equation}
where $(\phi_1,\phi_2)$ is a couple of two real independent solutions 
of the Schr\"odinger equation (SE), 
$-\hbar^2 \phi''/ 2m + V \phi = E \phi$, 
and $(a,b,\lambda)$ are real integration constants satisfying 
the condition $a \not= 0$, we obtain a first order differential 
equation representing the quantum law of motion \cite{BD3}.

Let us now examine the different formulations used to derive relation (5).
In \cite{BD1}, we can easily see that (5) is obtained from the 
Lagrangian  
\begin {equation}
L(x,\dot{x},a,b)={1\over 2} m {\dot{x}}^2 
                 \left( {\partial \hat{x}\over \partial x } \right)^2 
                 -V(x) \; .
\end {equation}
The non-classical integration constants $a$ and $b$, contained in  
$\hat{x}$, played the role of hidden parameters. Taking into account 
relation (3), $L$ and then the resulting quantum law of motion, 
depend also on the energy $E$. This constant is appeared also after 
the integration of the obtained equation of motion.
It follows that the integration constants are not independent and 
some of them form a redundant subset \cite{Floyd}. Although, this 
problem is gone round in \cite{BD2} by using in the Lagrangian 
formulation the coordinate $\hat{x}$, $L=L(\hat{x},\dot{\hat{x}})$, 
the fact remains that this formulation is not coherent when we use 
the coordinate $x$.

An analogous remark can be made also in the Hamiltonian formulation. 
In fact, relation (5) is reproduced in \cite{BD2} by using the 
Hamiltonian
\begin {equation}
H={P^2\over 2m}{\left({\partial x\over
 \partial \hat{x}}\right)}^2 +V(x) \; ,
\end {equation}
where $P=\partial S_0 / \partial x$ is the conjugate momentum. 
When the canonical equation was applied in \cite{BD2}, 
it is not taken into account the dependence of the factor 
$(\partial x/ \partial \hat{x})^2$ on $E$ and consequently on 
$H$. This observation was pointed out in the relativistic case in
\cite{BH}. In order to remedy this weakness, let us rewrite (8) 
in the following form 
\begin {equation}
P=\sqrt{2m(H-V)} \ {\partial \hat{x} \over \partial x} \; .
\end {equation}
With the use of (3) and (6), we can express 
$\partial \hat{x} /\partial x$
and by applying then (9) we can deduce $P$. Finally, the quantum 
law of motion can be derived from the canonical equation
\begin {equation}
\dot{x}= {\partial H \over \partial P} 
       = \left( {\partial P \over \partial H} \right)^{-1} \; .
\end {equation}
This equation leads to the same trajectories as those obtained 
from Jacobi's theorem,
$
t-t_0= {\partial S_0 / \partial E}  \; ,
$
as used by Floyd \cite{Fl1}. In fact, taking the derivative with 
respect to $x$ of this last relation, we obtain
\begin {equation}
{dt \over dx} = {\partial^2 S_0 \over \partial x \partial E}  
              = {\partial^2 S_0 \over \partial E \partial x}
              = {\partial P \over \partial E} 
              = {\partial P \over \partial H}    \; ,
\end {equation}
which is equivalent to (10). Consequently, when we 
take into account the dependence on $E$ of $\hat{x}$, as in Floyd's 
approach, the resulting trajectories will depend on the choice  
of the solutions of the SE \cite{BD2} that we will use in the 
reduced action in order to express ${\partial \hat{x} / \partial x}$.
For the moment, the only way to obtain a coherent Hamiltonian 
formulation leading to (5) is to use the coordinate system 
$(\hat{x})$ and to define the conjugate momentum as 
$\hat{P}={\partial \hat{S}_0 / \partial \hat{x}}$.
The transformation to the  coordinate system $(x)$ must not be performed 
until the canonical equation is applied.

The last observation that we make about relation (5) is pointed 
out in \cite{B2} and concerns turning points. At the points where 
$E=V(x)$, since $\partial S_0 / \partial x$ never has a vanishing 
value, we can show from (5) that all the higher temporal derivatives 
of $x$ take a vanishing value: $\dot{x}=0,\ddot{x}=0,...$ Thus, 
when the particle gets to one of these points, it can never leave 
it.

At first glance, the above three observations plead for a new 
law of motion. That's what is done in \cite{B2} where it is 
proposed a new approach consisting in the construction of a 
Lagrangian from which we have reproduced the QHJE. This approach was 
based on the following sensible hypothesis:
\begin{enumerate}
 
\item The Hamilton's principal function is an integral of a Lagrangian.

\item The Lagrangian is a difference between a kinetic term depending on 
($x$, $\dot{x}$, $\ddot{x}$, $\dot{\ddot{x}}$) and containing the 
quantum potential, and an external potential.

\item The resulting equation of motion is a fourth order one, in 
accordance with the one-dimensional QHJE.

\end{enumerate}
\noindent
In the context of these hypothesis, it is shown that in order to reach 
the QHJE from the constructed Lagrangian, the condition
\begin {equation}
{\partial S_0 \over \partial x} = m\dot{x} \; ,
\end {equation}
recalling the Bohm relation, is required. Relation (12) represented the 
new law of motion and it allowed with the use of (2) to establish 
the modified quantum Newton law \cite{B2}. Although, (12) 
reminds us of the classical mechanics, we stress that it describes 
the quantum motion because $\partial S_0 / \partial x$ represents the 
solution of the QHJE, Eq. (2). At first glance, this model 
seems attractive. However, as we will see in Section 4, Eq. (12) 
leads to a deadlock when we apply its two-dimensional version to the
motion of a charge submitted to an electric central potential.

\section{Extension to Higher Dimensions}

Although relation (12) has the same form as Bohm's law of motion, 
it is fundamentally different. In fact, in contrast to Bohm's theory,  
the reduced action $S_0$ in (12) is related to the wave function 
by (1) and therefore it never takes a constant value even in the case
where the wave function is real, up to a constant phase factor. 
As in Bohm's theory, the extension of relation (12) to three 
dimensions can be sensibly assumed as
\begin{equation}
m\dot{x}={\partial{S_{0}}\over\partial{x}} \; , \hskip20pt m\dot{y}=
{\partial{S_{0}}\over\partial{y}} \; ,\hskip20pt m\dot{z}=
{\partial{S_{0}}\over\partial{z}} \; .
\end{equation}

Concerning relation (5), before we extend it to higher dimensions, it is 
instructive to reproduce it in one dimension with a novel approach which 
we will use to perform this extension.

As $\hat{S}_0(\hat{x})= S_0(x)$, $\hat{V}(\hat{x})= V(x)$ 
and $\hat{E}= E$ \cite{FM2,BD1,FM4}, relation (4) can be written as 
\begin {equation}
{1\over 2m}{\left(\partial S_0(x) \over \partial x \right)}^2 
\left({\partial x \over \partial \hat{x}}\right)^2 +
V(x)=E  \; .
\end {equation}
On the other hand, if we use the coordinate system $(\hat{x})$ in 
which the quantum potential is canceled, the conjugate momentum takes 
the classical form
\begin {equation}
{\partial \hat{S}_0 \over \partial \hat{x}} = 
       m\dot{\hat{x}}  \; ,
\end {equation}
from which we deduce that
\begin {equation}
{\partial S_0 \over \partial x} 
\left({\partial x \over \partial \hat{x}} \right)^2 = 
       m\dot{x} \; .
\end {equation}
Substituting this expression in (14), we straightforwardly get 
to relation (5).

In what follows, we will use Einstein's convention for repeated 
indexes. Let us consider in a $D$-dimensional space the QHJE
\begin {equation}
\delta^{ij}{1\over 2m} 
                {\partial S_0 \over \partial x^{i} } 
                {\partial S_0 \over \partial x^{j} } -
                \delta^{ij}{\hbar^2 \over 2mR} 
                {\partial^2 R \over \partial x^{i}\partial x^{j}} + 
                V(x^i)=E  \; ,
\end {equation}
where the functions $R$ and $S_0$ satisfy the continuity equation
\begin {equation}
\delta^{ij} {\partial \over \partial x^i }
            \left(R^2 {\partial S_0 \over \partial x^j } \right) 
            =0    \; ,
\end {equation}
and $\delta^{ij}$ is the Kronecker symbol. In the coordinate system 
$(\hat{x}^i)$ in which the quantum potential is canceled, the QHJE 
takes the classical form
\begin {equation}
\delta^{ij}{1\over 2m} 
            {\partial \hat{S}_0 \over \partial \hat{x}^{i} } 
            {\partial \hat{S}_0 \over \partial \hat{x}^{j} } 
            + \hat{V}(\hat{x}^i)=\hat{E}  \; .
\end {equation}
As $\hat{S}_0(\hat{x}^i)= S_0(x^i)$, $\hat{V}(\hat{x}^i)= V(x^i)$ 
and $\hat{E}= E$, relation (19) turns out to be 
\begin {equation}
\delta^{ij}{1\over 2m}
    {\partial S_0 \over \partial x^l } 
    {\partial x^l \over \partial \hat{x}^i}
    {\partial S_0 \over \partial x^k }
    {\partial x^k \over \partial \hat{x}^j} +
    V(x^i)=E  \; .
\end {equation}
As in general relativity, we assume that the coordinate system
$(\hat{x}^i)$, in which the laws of motion take 
classical forms, is locally flat. Therefore, we have
\begin {equation}
{\partial \hat{S}_0 \over \partial \hat{x}^i} 
       = m{d \hat{x}_i \over dt} 
       = m \delta_{il} {d \hat{x}^l \over dt }\; ,
\end {equation}
from which we deduce that
\begin {equation}
{\partial S_0 \over \partial x^l} {\partial x^l \over \partial \hat{x}^i}  
   =  m \delta_{il} {\partial \hat{x}^l \over \partial x^n} \dot{x}^n \; .
\end {equation}
Multiplying each side of this last relation by
$\delta^{ij} \partial x^k / \partial \hat{x}^j$, we obtain
\begin {eqnarray}
\delta^{ij} {\partial S_0 \over \partial x^l} 
{\partial x^l \over \partial \hat{x}^i}   
{\partial x^k \over \partial \hat{x}^j} 
      & = & m \delta^{ij} \delta_{il} {\partial x^k \over \partial \hat{x}^j} 
              {\partial \hat{x}^l \over \partial x^n} 
            \dot{x}^n \nonumber\\
      & = & m \delta^{j}_{l}{\partial x^k \over \partial \hat{x}^j} 
              {\partial \hat{x}^l \over \partial x^n} 
            \dot{x}^n  \nonumber\\
      & = &  m {\partial x^k \over \partial \hat{x}^j} 
              {\partial \hat{x}^j \over \partial x^n} 
            \dot{x}^n   \nonumber\\  
      & = &  m {\partial x^k \over \partial x^n} \dot{x}^n   \nonumber\\ 
      & = &  m \delta^{k}_{n} \dot{x}^n   \nonumber\\  
      & = & m \dot{x}^k \; .
\end {eqnarray}
Using this result in (20), we find
\begin {equation}
{1\over 2}{\partial {S}_0 \over \partial x^k } \; \dot{x}^k +V(x^i)=E \; .
\end {equation}
This represents the higher dimension version of relation (5).  Although, 
relation (24) works in classical mechanics ($\partial S^{clas}_0 / 
\partial x^k = m\dot{x}_k $), it describes the quantum motion 
because in (24) $\partial S_0 / \partial x^k$ is the solution in 
higher dimensions of the QHJE, already investigated in \cite{BM}. 
The problem of the immobility of particles at turning points 
disappears. In fact, when the space dimension is higher than 
one, we cannot show from (24) that the derivatives 
$\dot{x}^k, \ddot{x}^k,...$ ($k=1,2,..., D$) take simultaneously 
vanishing values.

\section{Hydrogen atom in two dimensions}

In this Section, let us apply in two dimensions the laws of motion 
(13) and (24) to the hydrogen atom for which the potential is 
\begin {equation}
V(r) = -{e^2\over r} \; ,
\end {equation}
where $e^2=q^2/4\pi\epsilon_0$, $q$ being the absolute value of the 
electron charge. By using polar coordinates 
$(r,\theta)$ and writing $\psi(r,\theta)=R(r)\Theta(\theta)$, 
it is well known that Schr\"odinger's equation
\begin {equation}
-{\hbar^2\over 2m}\Delta\psi+V(r)\psi=E\psi \; ,
\end {equation}
leads to the two following separated relations
\begin {equation}
{d^2R\over d\rho^2}+{1\over \rho}{dR\over d\rho}+
\left[
    {2\over \rho}-{l^2\over \rho^2}-\alpha^2
   \right]R
   =0 \; ,
\end {equation}
and
\begin {equation}
{d^2\Theta\over d\theta^2}+l^2\Theta=0 \; ,
\end {equation}
where $l$ is an integration constant, $\rho=r/a_0$ and $\alpha^2=-E/E_I$, 
$a_0$ and $E_I$ being respectively Bohr's atomic radius $(a_0=\hbar^2/me^2)$ 
and the ionization energy $(E_I=me^4/2\hbar^2)$ of the Hydrogen atom. 
Choosing for (28) as independent real solutions the two following functions
\begin {equation}
\Theta_1=\cos l\theta \ , \ \ \ \ \ \ \ \ \ \ \ \ \Theta_2=\sin l\theta \; , 
\end {equation}
and imposing the conditions
\begin {equation}
\Theta_1(\theta)=\Theta_1(\theta +2\pi) \ , \ \ \ \ \ \ \ \ \ \ \ \ \
\Theta_2(\theta)=\Theta_2(\theta +2\pi) \ , 
\end {equation}
we deduce that $l$ must be an integer number. In \cite{PP}, 
it is shown that a physical solution for (27) is 
\begin {equation}
R_1(\rho)=\rho^{|l|}\exp(-\alpha\rho)L_{n-|l|}^{2|l|}(2\alpha\rho)  \; ,
\end {equation}
where $L_{s}^{k}$ are the generalized Laguerre polynomials, 
$n=0,1,2,3 \ldots$ is the principal quantum number 
$(-n \leq l \leq n)$ and 
\begin {equation}
\alpha=\frac{1}{n+1/2} \; .
\end {equation}
This last relation leads to 
\begin {equation}
E(n)\equiv E_n =-\frac{E_I}{(n+1/2)^2} \; .
\end {equation}
A second real independent solution $R_2(\rho)$ for (27) can be derived by 
using the Wronskian $W(R_1,R_2)$ \cite{Pisk}
\begin {equation}
R_2(\rho)= R_1(\rho) \int\frac{\exp \left[-\int\frac{d\rho}{\rho}\right]}
           {R_1^2(\rho)} \ d\rho  
         = R_1(\rho) \int\frac{d\rho}{\rho R_1^2(\rho)}  \; .
\end {equation} 
As shown in \cite{BM}, the reduced action in two dimensions is 
\begin {equation}
S_0= \hbar \ \arctan \left[
                 \frac
                    {R_1\Theta_1 + \nu_2 R_1\Theta_2
                 + \nu_3 R_2 \Theta_1 + \nu_4 R_2\Theta_2}
                    {\mu_1 R_1\Theta_1 + \mu_2 R_1 \Theta_2
                 + \mu_3 R_2 \Theta_1 +  R_2\Theta_2}
             \right]  
                 + \hbar \lambda \; ,
\end {equation} 
where $(\nu_2, \nu_3, \nu_4, \mu_1, \mu_2, \mu_3, \lambda) $ are real 
integration constants. For the ground state $(n=0, l=0)$, by (32) we have
$\alpha=2$. As $L_{0}^{0}=1$ and $\rho=r/a_0$, with the use of (29), (31) 
and (34), we deduce that 
\begin {equation}
\Theta_1=1 \ , \ \ \ \ \ \ \ \ \ \ \ \ \Theta_2=0 \; ,
\end {equation}
and
\begin {equation}
R_1 = \exp \left(-\frac {2r}{a_0} \right) \; , 
\ \ \ \ \ \ \ \ \ \ \ \   R_2= \exp \left( - \frac {2r}{a_0} \right) \
\int^{r}_{r_0} \frac{\exp (4r'/a_0)}{r'} \ dr'  \; ,       
\end {equation}
Note that the lower boundary $r_0$ can be arbitrary chosen. Therefore, 
in order to avoid the singular point $r'=0$ in (37), we choose $r_0$ 
positive. Note also that with suitable integration boundaries, the 
integral in (37) can be identified to the exponential integral 
\textbf{Ei}(r) defined as the Cauchy's principal value of
\begin {equation}
\textbf{Ei}(x)=\int^{x}_{-\infty} \frac{\exp (t)}{t} dt 
\ \ \ \ \ \ \ \ \ \ (x>0) \; .       
\end {equation}
It follows that for the ground state, expression (35) reduces to
\begin {equation}
S_0= \hbar \ \arctan \left[
                 \frac
                    {R_1 + \nu_3 R_2}
                     {\mu_1 R_1 + \mu_3 R_2}
             \right]  
                 + \hbar \lambda \; ,
\end {equation} 
where $R_1(r)$ and $R_2(r)$ are given in (37). 

In polar coordinates, the law of motion (13) takes the form
\begin {equation}
 \frac{\partial S_0}{\partial r} = m \dot{r}\ , \ \ \ \ \ \ \ \ \ \ \ \ 
         \frac{\partial S_0}{\partial \theta} =mr^2\dot{\theta} \; ,
\end {equation} 
while the law of motion (24) turns out to be
\begin {equation}
\dot{r} \frac{\partial S_0}{\partial r}
        + \dot{\theta} \frac{\partial S_0}{\partial \theta}
                                  =2\left[E-V(r) \right] \; .
\end {equation} 
Substituting (39) in (40), we get to
\begin {eqnarray}
\hbar \left( \mu_1\nu_3 - \mu_3\right) 
\left[ \frac{2}{a_0} \exp \left(-\frac{4r}{a_0} \right)
        \int^{r}_{r_0} \frac{\exp (4r'/a_0)}{r'} \ dr' \right. + 
        \hskip26mm&& \nonumber \\
        \left. \frac{1}{r}\exp \left( \frac {2r}{a_0} \right) 
                 \right] 
                                  = m\dot{r}H(r) \; ,
\end {eqnarray}
and
\begin {equation} 
mr^2\dot{\theta}=0 \; ,
\end {equation} 
where   
$H(r)=\left(R_1 +  \nu_3 R_2\right)^2+ 
\left(\mu_1 R_1 +  \mu_3 R_2\right)^2$.
However, with the use of (25) and (33), and by substituting 
(39) in (41), we get to
\begin {eqnarray}
\hbar \left( \mu_1\nu_3 - \mu_3 \right) \dot{r} 
\left[ \frac{2}{a_0} \exp \left(-\frac{4r}{a_0} \right)
        \int^{r}_{r_0} \frac{\exp (4r'/a_0)}{r'} \ dr' 
        \right. + \hskip26mm&& \nonumber \\
        \left. \frac{1}{r}\exp \left( \frac {2r}{a_0} \right) 
                 \right] 
                          =2 H(r) \left[-\frac{2me^4}{\hbar^2}
                                           +\frac{e^2}{r} \right] \; .
\end {eqnarray} 
Relation (43) results from (13) and indicates that $\theta$ is constant. 
This result is unacceptable and forced us to abandon the law of motion (13).   
Relation (44) results from (24) and does not allow to determine 
$\theta$. This indicates that relation (24) does not ensure a complete 
description of the quantum motion when the number of degrees of freedom is 
higher than one. However, this is not sufficient to reject relation (24). 
In fact, even in classical mechanics, it is well-known that the  
law of energy conservation does not allow to determine completely 
the motion except for systems of one degree of freedom.
This feature encourage us to pay  attention to relation (5) and 
its extended version (24).

\section{ The Relativistic Case }

Before we establish the relativistic version of (24), it is instructive  
to remind us of the following main points:
\begin{enumerate}

\item  When the coordinate system $(\hat{x})$ is used to apply Jacobi's 
theorem \cite{BD1} or to express the Lagrangian in order to obtain 
the equation of motion \cite{BD2}, relation (5) is derived without 
any mathematical ambiguity. It is also the case for the Hamiltonian 
formulation.

\item In the coordinate system $(\hat{x})$, 
the QHJE and the other motion laws take the classical forms
meaning that the choice of this system is made in such a way  as to 
cancel the effect of the quantum potential. This strongly reminds us 
of the equivalence principle of general relativity which allows to admit 
the existence of a coordinate system in which the gravitational 
field is locally canceled.

\item  Compared to the CHJE, the quantum potential is an additional term 
which guarantees the covariance of the QHJE \cite{FM2}. The same role is 
also played by the gravitational field in general relativity.

\item The gravitational potential is generated by the quantum 
potential \cite{M}.

\end{enumerate}
Unquestionably, the above remarks plead in favor of (5) and (24). 
Consequently, if we would like to continue to believe in the link
between the quantum potential and the gravitational field, we have
a further argument to abandon the quantum law (12) and also the hypothesis, 
though attractive, enumerated in Section 2 and which 
allowed in \cite{B2} to establish (12).
Concerning the reservations expressed in Section 2 about (5), we 
can make the following observations. The absence for the moment of 
a coherent Lagrangian or Hamiltonian formulation in any system coordinate, 
except in $(\hat{x})$ in which the quantum potential is canceled, 
must not imply the rejection of the approach. With regard to the particle 
immobility at turning points, pointed out in \cite{B2},
this problem appears only in one dimension. As we have seen in Section 3, 
this problem disappears in realistic models for which the space dimension 
is higher than one. 

Let us now establish the relativistic version of (24). In the context of 
the equivalence postulate, the $D$-dimensional 
stationary relativistic QHJE for spinless system reads \cite{BFM}
\begin {equation}
\delta^{ij}{1\over 2m} 
                {\partial S_0 \over \partial x^{i} } 
                {\partial S_0 \over \partial x^{j} } -
                \delta^{ij}{\hbar^2 \over 2mR} 
                {\partial^2 R \over \partial x^{i}\partial x^{j}} + 
                {m^2 c^4-[E-V(x^i)]^2 \over 2mc^2}=0  \; ,
\end {equation}
where the functions $R$ and $S_0$ satisfy the continuity equation
\begin {equation}
\delta^{ij} {\partial \over \partial x^i }
            \left(R^2 {\partial S_0 \over \partial x^j } \right) 
            =0    \; .
\end {equation}
The summation on $i$ and $j$ does not concern the time component 
\cite{FM2}. In the coordinate system $(\hat{x}^i)$ in which the quantum 
potential is canceled, the relativistic QHJE takes the classical form
\begin {equation}
\delta^{ij}{1\over 2m} 
            {\partial \hat{S}_0 \over \partial \hat{x}^{i} } 
            {\partial \hat{S}_0 \over \partial \hat{x}^{j} } 
            + {m^2 c^4-[\hat{E}-\hat{V}(\hat{x}^i)]^2 \over 2mc^2}=0  \; .
\end {equation}
As $\hat{S}_0(\hat{x}^i)= S_0(x^i)$, $\hat{V}(\hat{x}^i)= V(x^i)$ 
and $\hat{E}= E$, relation (47) turns out to be 
\begin {equation}
\delta^{ij}{1\over 2m}
    {\partial S_0 \over \partial x^l } 
    {\partial x^l \over \partial \hat{x}^i}
    {\partial S_0 \over \partial x^k }
    {\partial x^k \over \partial \hat{x}^j} +
    {m^2 c^4-[E-V(x^i)]^2 \over 2mc^2}=0   \; .
\end {equation}
On the other hand, the coordinate system $(\hat{x}^i)$ being locally 
flat, the conjugate momentum takes the following classical relativistic form
\begin {equation}
{\partial \hat{S}_0 \over \partial \hat{x}^i} 
       =  m{d \hat{x}_i \over d \tau } 
       = m \delta_{il} {d \hat{x}^l \over d \tau } \; ,
\end {equation}
where $d\tau$ is an element of the proper time associated to the 
particle. Relation (49) can be obtained from (21) by substituting $dt$ 
by $d\tau$. Thus, in the same manner as in Section 3, we can show 
from (49) that
\begin {equation}
\delta^{ij} {\partial S_0 \over \partial x^l} 
{\partial x^l \over \partial \hat{x}^i}   
{\partial x^k \over \partial \hat{x}^j} 
                                          =  m {dx^k \over d \tau }\; .
\end {equation}
Using this result in (48), we find
\begin {equation}
{\partial {S}_0 \over \partial x^k } {dx^k \over d\tau} 
           + {m^2 c^4-[E-V(x^i)]^2 \over mc^2}   =0 \; .
\end {equation}
This represents the relativistic quantum law of motion in  
higher dimensions. Although, relation (51) works in classical 
relativistic mechanics 
($\partial S^{clas/relat}_0 / \partial x^k $   $=  m dx_k/d\tau $), 
it describes the relativistic quantum motion because in (51) 
$\partial S_0 / \partial x^k$ represents the solution of the 
relativistic QHJE, Eq. (45). As in the non-relativistic case 
\cite{BM}, one can check that the solutions of (45) and (46) are 
\begin{equation}
S_0=\hbar \ \arctan{\left(\phi_1\over\phi_2\right)}
+\hbar l \; ,
\end{equation}
and
\begin{equation}
R=k \sqrt{{\phi_1^2}+{\phi_2^2}} \; ,
\end{equation}
where $\phi_1$ and $\phi_2$ are two real independent solutions 
of the stationary Klein-Gordon equation,
\begin{equation}
-{\hbar^2 \over 2m} \Delta \phi 
           + {m^2 c^4-[E-V(x^i)]^2 \over 2mc^2} \phi = 0 \; ,
\end{equation}
$l$ and $k$ arbitrary integration constants, and $\Delta$ the 
$D$-dimensional Laplacian. We mention that in the separated 
variable case, it is possible to make explicit in (52) all the 
integration constants, as in the non-relativistic case \cite{BM}. 

Finally, we indicate that in one dimension, relation (51) 
reproduces the same result as the one obtained in \cite{BH}. In fact,
(51) allows us to write 
\begin {equation}
{\partial {S}_0 \over \partial x } {dx \over d\tau} 
           + {m^2 c^4-[E-V(x)]^2 \over mc^2}   =0 \; .
\end {equation}
Multiplying by $d\tau/dt$, we have
\begin {equation}
{\partial {S}_0 \over \partial x } {dx \over dt} 
           + {m^2 c^4-[E-V(x)]^2 \over mc^2}{d\tau \over dt}   =0 \; .
\end {equation}
On the other hand, the time component being not concerned by the 
coordinate transformation $(\hat{t}=t)$ \cite{FM2}, in the 
system $(\hat{x})$ where the space is locally flat, we have
\begin {equation}
d\tau^2=dt^2-{1 \over c^2}d\hat{x}^2
       = dt^2-{1 \over c^2} \left({\partial \hat{x} 
         \over \partial x}\right)^2 dx^2 \; .
\end {equation}
By substituting in this last relation the expression
\begin {equation}
\left({\partial\hat{x}  \over \partial x }\right)^2 
                      = c^2{(\partial S_0/ \partial x)^2 
                                      \over [E-V(x)]^2-m^2c^4} \; ,
\end {equation}
which defines the transformation $x \to \hat{x}$ allowing 
to cancel the quantum potential \cite{BH}, we get to
\begin {equation}
\left({d\tau \over dt }\right)^2 
                      = {[E-V(x)]^2-m^2c^4 - 
                        \dot{x}^2(\partial S_0/ \partial x)^2 
                                      \over [E-V(x)]^2-m^2c^4 } \; .
\end {equation}
Using this expression in (56), we reproduce the one-dimensional 
relativistic quantum law of motion,
\begin {equation}
{\partial {S}_0 \over \partial x } {dx \over dt} 
           + {m^2 c^4-[E-V(x)]^2 \over E-V(x)}   =0 \; ,
\end {equation}
already obtained in \cite{BH}.

\section{ Conclusion }

Before summarizing the main results of the present investigation, 
let us come back to the quantum law (5) established in 
\cite{BD1}. Despite its present insufficiencies observed in Section 2, 
we think that it is useful to investigate this law and its extended 
versions in more detail for the following reasons.

\begin{enumerate}

\item The problem of the particle immobility at turning points is 
specific to the one-dimensional space. In a realistic model, we cannot 
ignore the other dimensions of the space and this 
problem disappears.

\item The fact of the absence for the moment of a coherent 
Lagrangian or Hamiltonian formulation which works in any coordinate 
system does not mean that law (5) must be rejected. In fact, in the 
same manner as for the law of motion in general relativity, relation 
(5) is rigorously established in the particular system $(\hat{x})$ 
in which the quantum potential is canceled. The transformation 
from $(\hat{x})$ to another system $(x)$, which can be for example 
the laboratory frame, is performed after the equation of motion 
is obtained in the system $(\hat{x})$. 

\item Another interesting feature of (5) is the nodal structure of the 
quantum trajectories which follow from it. In fact, it is shown in 
\cite{BD3} that to each classical trajectory there is a family of 
quantum trajectories which all pass through some points 
constituting nodes and belonging to the classical trajectory.  
In addition, there is an interesting relation between de Broglie's 
wavelength and the length separating adjacent nodes which become 
infinitely close in the limit $\hbar \to 0$. Also, it is shown 
\cite{BD3} that in the classical limit all the quantum trajectories 
tend to be identical to the classical one.

\end{enumerate}

Furthermore, the manner in which the quantum law (5) is derived 
allows to establish a parallel between the postulate equivalence 
of quantum mechanics and the one of general relativity. 
This suggests that relation (5) and its extended versions may 
play an important role in the search for a possible link, 
already investigated by Matone \cite{M} and Carroll \cite{Ca}, 
between the quantum potential and the gravitational potential.

To summarize, it is in this spirit that we have performed in Sections 
3 and 5 the extension to higher dimensions of relation (5) respectively 
in the non-relativistic case and the relativistic one. We have also 
applied its two-dimensional version in the non-relativistic case to 
the hydrogen atom. An interesting question is how to complete 
relations (24) and (51) in order to describe in its entirety the quantum 
motion in any dimension.

\bigskip
\bigskip

{\bf References}

\end {document}